\documentclass[a4paper,11pt]{article}
\usepackage{jheppub}
\usepackage{multicol,enumerate}
\usepackage{multirow,amsmath,amssymb}
\usepackage{subfigure}
\usepackage{slashed} 
\usepackage{color} 

\newcommand{\fbinv} {\mbox{\ensuremath{~\text{fb}^\text{$-$1}}}}

\newcommand{\pythia}{{\sc Pythia}}

\newcommand{\whizard}{{\sc Whizard}}
\newcommand{\madgraph}{{\sc MadGraph}}

\def\Hboson{\ensuremath{H}}%
\def\ee{\ensuremath{e^+ e^-}}%
\def\mm{\ensuremath{\mu^+ \mu^-}}%
\def\tt{\ensuremath{\tau^+ \tau^-}}%
\def\Zboson{\ensuremath{Z}}

\title{$\Hboson \xrightarrow {}\ee$ at CEPC: ISR effect with MadGraph}

\author{Cheng Chen $^a$,}
\author{Zhenwei Cui $^a$,}
\author{Gang Li$^{b,c}$,}
\author{Qiang Li$^{a,b}$,}
\author{Manqi Ruan$^{b,c}$,}
\author{Lei Wang$^a$,}
\author{Qi-shu Yan$^d$}

\affiliation{$^a$Department of Physics and State Key Laboratory of Nuclear Physics and Technology, \\
Peking University, Beijing, 100871, China}
\affiliation{$^b$CAS Center for Excellence in Particle Physics, Beijing 100049, China}
\affiliation{$^c$ Institute of High Energy Physics, Beijing 100049, China}
\affiliation{$^d$College of Physics Sciences, University of Chinese Academy of Sciences, Beijing 100049, China and Center for High Energy Physics, Peking University, Beijing 100871, China}

\emailAdd{qliphy0@pku.edu.cn, li.gang@ihep.ac.cn, melodyphysics@gmail.com}

\abstract{The Circular Electron Positron Collider (CEPC) is a future Higgs factory proposed by the Chinese high energy physics community. It will operate at a center-of-mass energy of 240-250 GeV. The CEPC will accumulate an integrated luminosity of 5 ab$^{\rm{-1}}$ in ten years' operation. With GEANT4-based full simulation samples for CEPC, Higgs boson decaying into electron pair is studied at the CEPC. The upper limit of ${\cal B}(\Hboson \xrightarrow {}\ee)$ could reach 0.024\% at 95\% confidence level.  The signal process is generated by \madgraph , with Initial State Radiation (ISR) implemented~\footnote{\url{http://www.phy.pku.edu.cn/~qiangli/mgisr.html}}, as a first step to adjust \madgraph~ for a electron positron Collider. }

\date{\Date}


\begin{document}
\maketitle
\flushbottom

\section{Introduction}
\label{intr}

\qquad The amazing discovery of Higgs boson~\cite{ref:1,ref:2} in 2012 by the ATLAS and CMS experiments at the CERN LHC has made a considerable step in particle physics, opening doors to new physics search through Higgs portal.   The up-to-date results indicate that it is highly Standard Model (SM) like~\cite{cmshig,lhcsub1,lhcsub2,lhcsub3,lhcsub4,lhcsub5}. However, many new physics models predict the Higgs couplings deviate from the SM at the percent level. Thus the percent or even sub-percent level precision becomes necessary for the future Higgs measurement program. With this consideration, a Higgs factory at $e^+e^-$ collider with high luminosity is best suited for this goal, due to its clean environment and relative lower cost.

The Circular Electron-Positron Collider(CEPC)~\cite{ref:3} is such a nice example, which is a proposed circular collider, designed to run around $240\sim250$ GeV with an instantaneous luminosity of 2 $\times$ $\rm{10}^{\rm{34}}$ $\rm{cm}^{\rm{-2}}$ $\rm{s}^{\rm{-1}}$, and will deliver 5 $\mathrm{ab}^{-1}$ of integrated luminosity during 10 years of operation. About $10^6$ Higgs events will be produced in a clean environment,  which allows the measurement of the cross section of the Higgs production as well as its mass, decay width and branching ratios with precision much beyond those of hadron colliders.

\begin{figure}[!t]
  \centering
  \includegraphics[width=0.4\textwidth]{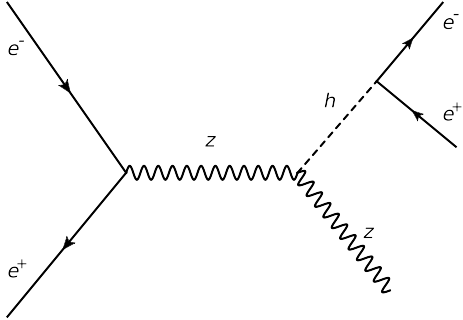}
  \caption{\label{fig:fmd} Example Feynman diagram.}
\end{figure}

At CEPC with the center-of-mass energy of 250 GeV, the Higgs bosons are dominantly produced from $ZH$ process, where the Higgs boson is produced in association with a $Z$ boson. Major deay modes of the Higgs boson have been extensively studied in Refs~\cite{ref:3,Chen:2016zpw}, such as the channel of  $\Hboson \xrightarrow {}\Zboson\Zboson$, and  $\Hboson \xrightarrow {}\gamma\gamma$ etc. In this study we are interested in a rare decay $\Hboson \xrightarrow {}\ee$. The Feynman diagram of $\Hboson \xrightarrow {}\ee$ is shown in Figure.~\ref{fig:fmd}. 

The SM prediction for the branching fraction ${\cal B}(\Hboson \xrightarrow {}\ee)$ is as tiny as approximately $5\times 10^{-9}$. However, in new physics scenario (see e.g.~\cite{Altmannshofer:2015qra}), it can be enhanced significantly. Moreover, searching or measurement for $\Hboson \xrightarrow {}\ee$ together with $\mm$ and $\tt$, can be used to test the lepton universality of Higgs boson couplings.  

The two electrons from Higgs decay can be easily identified and their momentum can be precisely measured in the detector. The Higgsstrahlung events can then be reconstructed with the recoil mass method:
\begin{equation}
m_{recoil}^{2}=s+m^{2}_{H}-2\cdot E_{H}\cdot~ \sqrt[]s ~,
\end{equation}
where $\sqrt[]s$ is the center of mass energy, $m_{H}$ and $E_{H}$ are the mass and energy of the Higgs boson reconstructed by the two lepton four momentum. Therefore, the $ZH$ ($\Hboson \xrightarrow {}\ee$) events form a peak in the $M_{\rm{recoil}}$ distribution at the Z boson mass. With the recoil mass method, the $ZH$ events are selected without using the decay information of the Z boson.

A search has already been performed at CMS with RunI data~\cite{Khachatryan:2014aep}, with an upper limit of 0.19\% placed on the branching fraction ${\cal B}(\Hboson \xrightarrow {}\ee)$.  Studies through resonant s-channel $\ee \xrightarrow {} \Hboson$ have also been proposed at FCC-ee~\cite{fccee} operating at a collison energy of 125 GeV, with sensitivies being able to reach down to 2 times SM prediction with  10 $\mathrm{ab}^{-1}$ of integrated luminosity, depending, however, on good controls on beam spread.

This paper is organized as follows. Section 2 shows the ISR implementation in \madgraph . Section 3 describes the detector model, Monte Carlo (MC) simulation and samples used in the studies.  Section 4 presents the measurements of $\Hboson \xrightarrow {}\ee$. The conclusion is summarized in Section 5.

\section{ISR implementation in MadGraph}
\label{isrmg}
\begin{figure}[!t]
  \centering
  \includegraphics[width=0.48\textwidth]{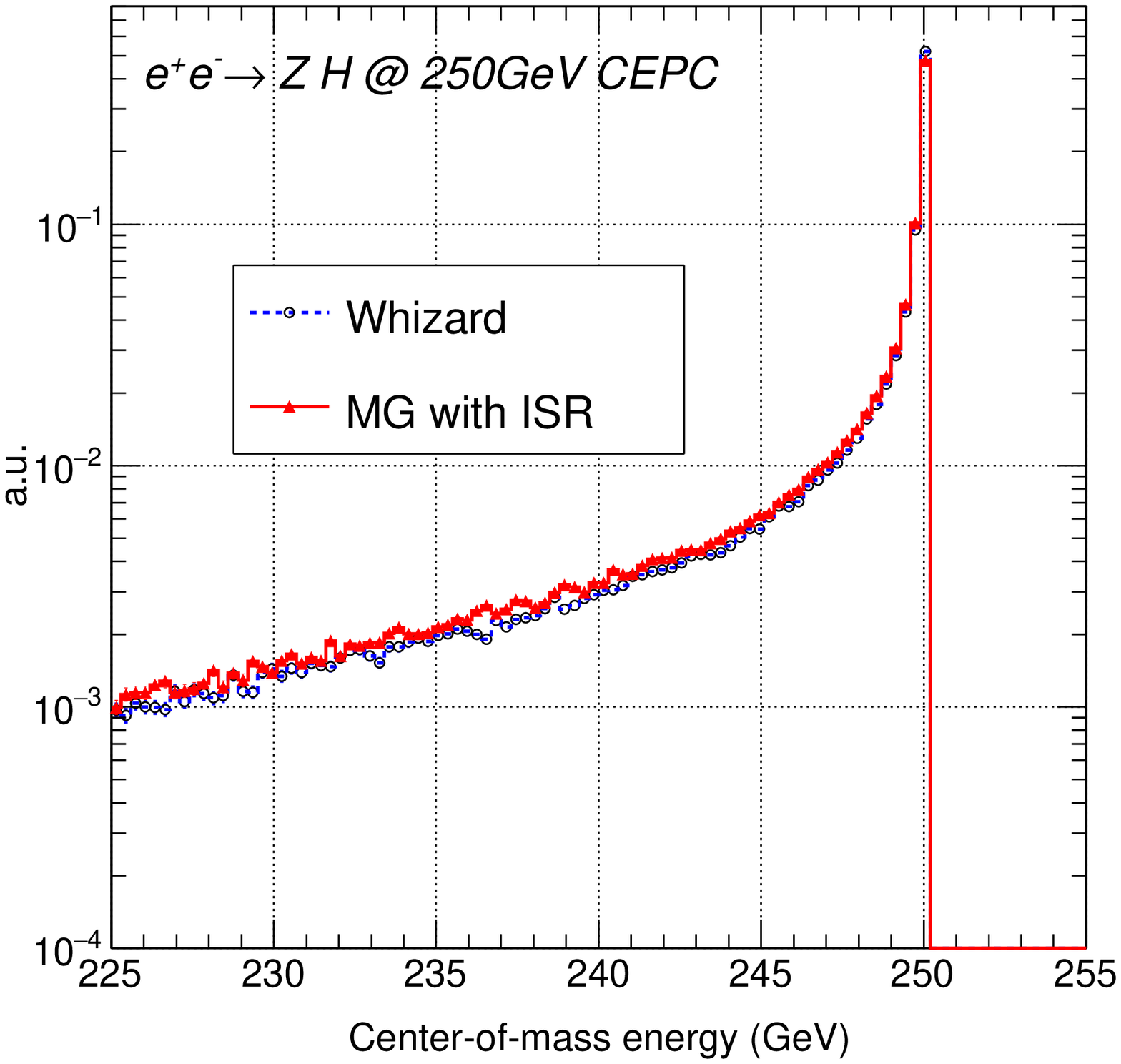}
  \includegraphics[width=0.48\textwidth]{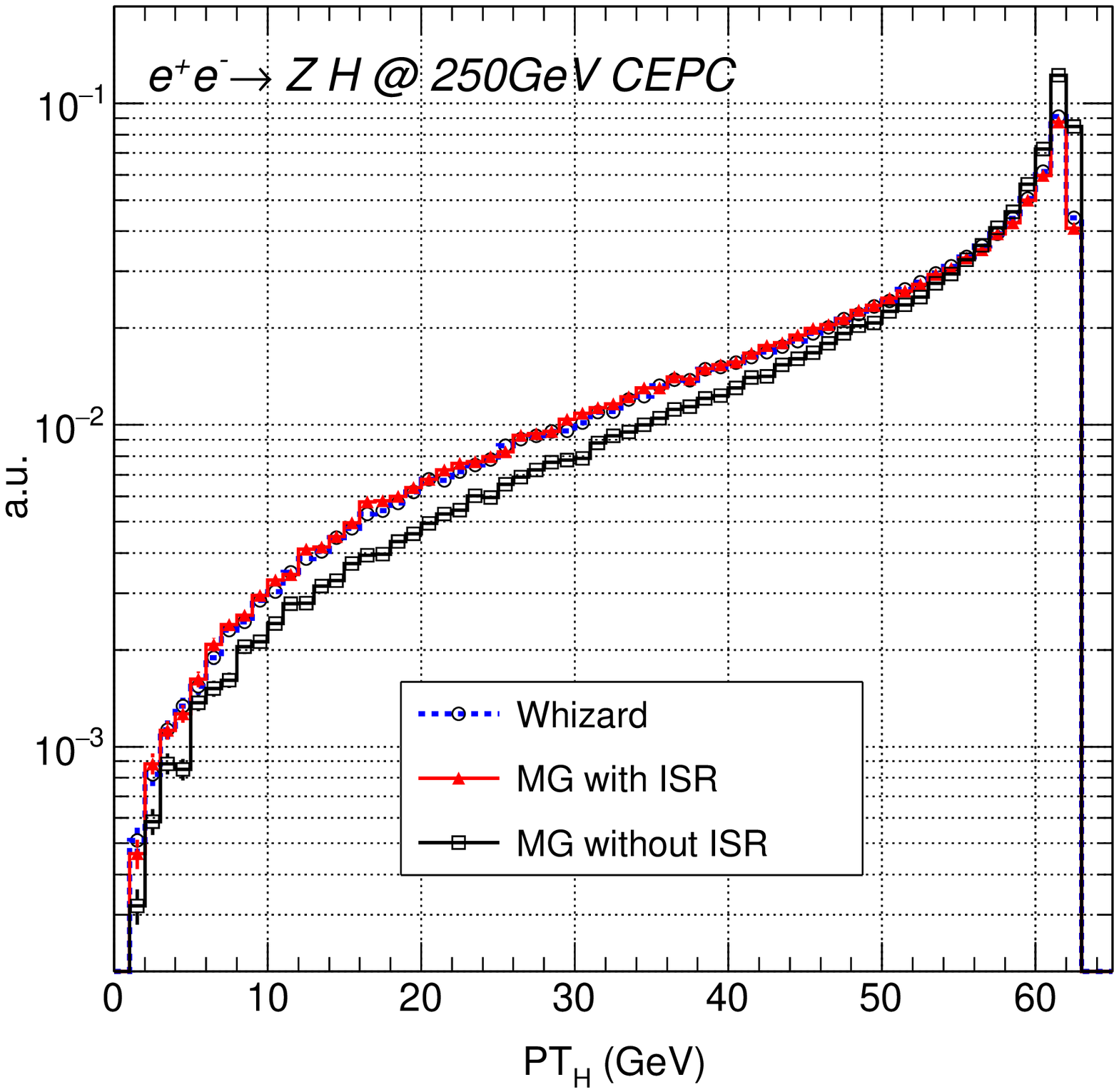}
  \caption{\label{fig:isr} Comparisons plots on center-of-mass energy and Higgs transverse momentum,  between \whizard~ and \madgraph~ with or without ISR effect included, for the process $e^+e^- \rightarrow ZH$.}
\end{figure}

\qquad The Initial State Radiation (ISR) is an important issue in high energy processes, especially for lepton colliders. ISR affects cross section significantly, for example, reduces the $ZH$ cross section by more than 10\%.  Following Whizard~\cite{ref:4}, we have implemented in \madgraph~ the lepton ISR structure function that includes all orders of soft and soft-collinear photons as well as up to the third order in hard-collinear photons. Comparisons can be seen in Fig.~\ref{fig:isr} for  $e^+e^- \rightarrow ZH$, from which one can see the good agreement between \whizard and \madgraph~ with ISR included, on distributions of center-of-mass energy and Higgs transverse momentum.  Similar checks have also been passed for other processes including for the process $e^+e^- \rightarrow W^+W^-$ and $W^+W^-Z$. 

One should note that besides ISR, another macro effect at high luminosity electron-positron collider, beamstrahlung, also affects the cross section.  In the storage ring the beamstrahlung  effect makes the beam energy spread larger and reduces the center of mass energy~\cite{ref:cepc_acc}. The effect, however, are found to be small at CEPC. 

Based on above progress, we are now able to generate signal samples in \madgraph~ with ISR effect included, for $e^+e^- \rightarrow ZH$, together with  the decay of $\Hboson \xrightarrow {}\ee$  at matrix element level, thanks to the convinience of \madgraph. 

\section{Detector and Simulation}
\label{isrmg}
\qquad The analysis is performed on the MC samples simulated on the CEPC conceptual detector, which is based on the International Large Detector (ILD)~\cite{ild1,ild2} at the ILC~\cite{ILC}. At CEPC, electron identification efficiency is expected to be over 99.5\% for $p_{\rm{T}}$ larger than 10 GeV, and with excellent $p_{\rm{T}}$ resolution of $\sigma_{1/p_{\rm{T}}} = 2\times 10^{-5} \oplus 1\times 10^{-3}/(p_{\rm{T}}\sin\theta)$. More details can be checked in~\cite{ref:3,Chen:2016zpw}.

For the signal process, $e^+e^- \rightarrow ZH$ with $\Hboson \xrightarrow {}\ee$, 50K events are  generated by \madgraph~ V2.3.3 with ISR effect included, with Higgs mass set to be 125 GeV. For the backgrounds, \whizard~ V2.2.8~\cite{ref:4}, are exploited as the event generator. All these samples are produced at the center-of-mass energy of 250 GeV. 

The major SM backgrounds,
including all the 2-fermion processes($e^+e^-\rightarrow f\bar{f}$,
where $f\bar{f}$ refers to all lepton and quark pairs except $t\bar{t}$) and 4-fermion processes($ZZ$, $WW$, $ZZ$ or $WW$, single $Z$, single $W$). The initial states radiation (ISR) and all possible interference effects are taken into account in the generation automatically. The classification for four fermions production, is referred to LEP~\cite{ref:17}, depending crucially on the final state.
For example, if the final states consist of two mutually charge conjugated fermion pairs that could decay from both $WW$ and $ZZ$ intermediate state, such as $e^{+}e^{-}\nu_{e}\bar{\nu_{e}}$,
this process is classified as ``$ZZ$ or $WW$'' process.
If there are $e^{\pm}$ together with its parter neutrino and an on-shell $W$ boson in the final state, this type is named as ``single $W$''.  Meanwhile, if there are a electron-positron pair and a on-shell $Z$ boson  in the final state, this case is named as ``single $Z$''.
More details about the CEPC samples set can be found in reference~\cite{Mo:2015mza}.

Signal and background samples are further interfaced with \pythia\ 6~\cite{pythia} for parton shower and hadronization, and then fully simulated with Mokka~\cite{Mokka} and reconstructed with ArborPFA~\cite{arbor}.

\section{Results}
\label{results}

\begin{figure}[!t]
  \centering
  \includegraphics[width=0.4\textwidth]{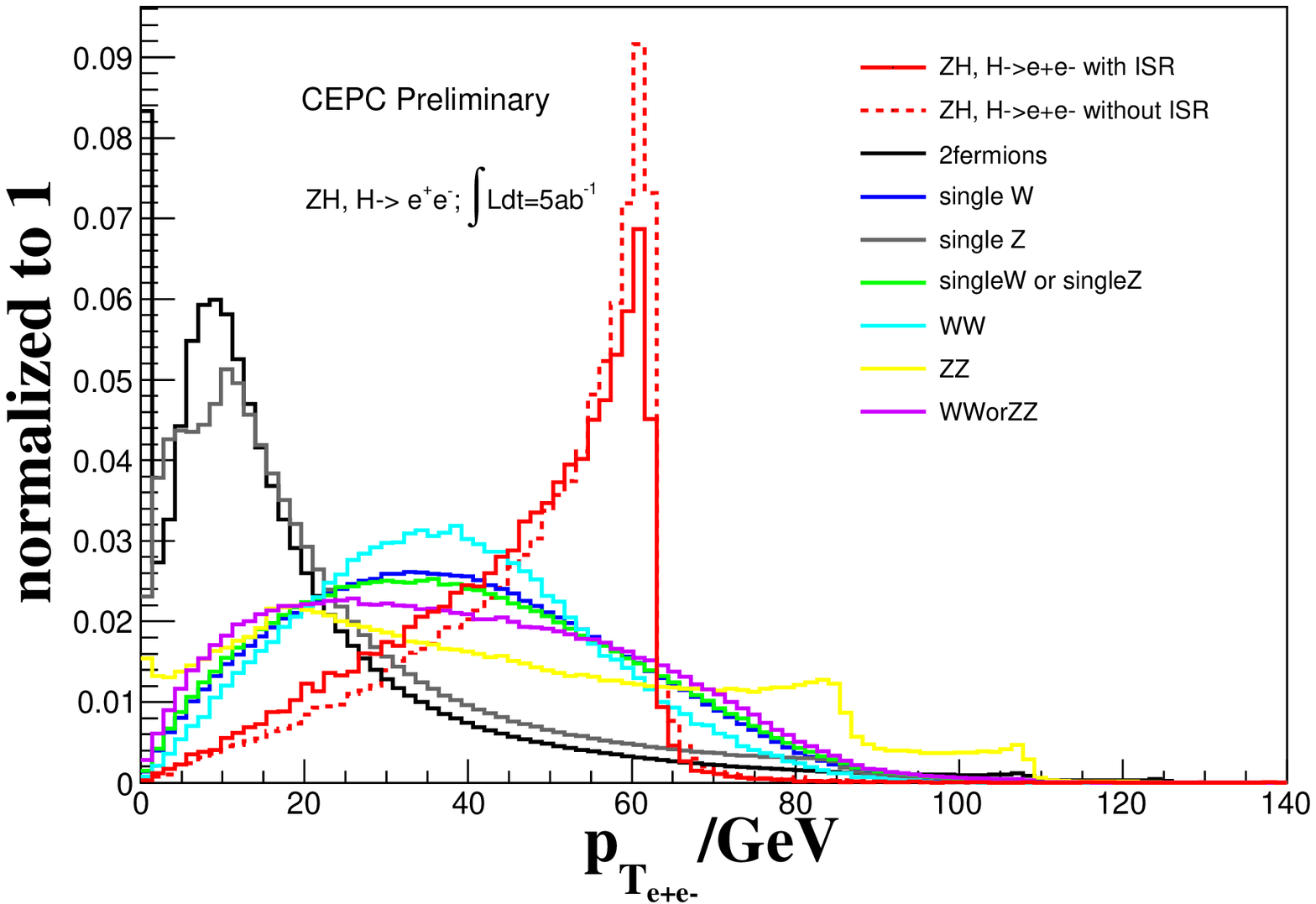}
  \includegraphics[width=0.4\textwidth]{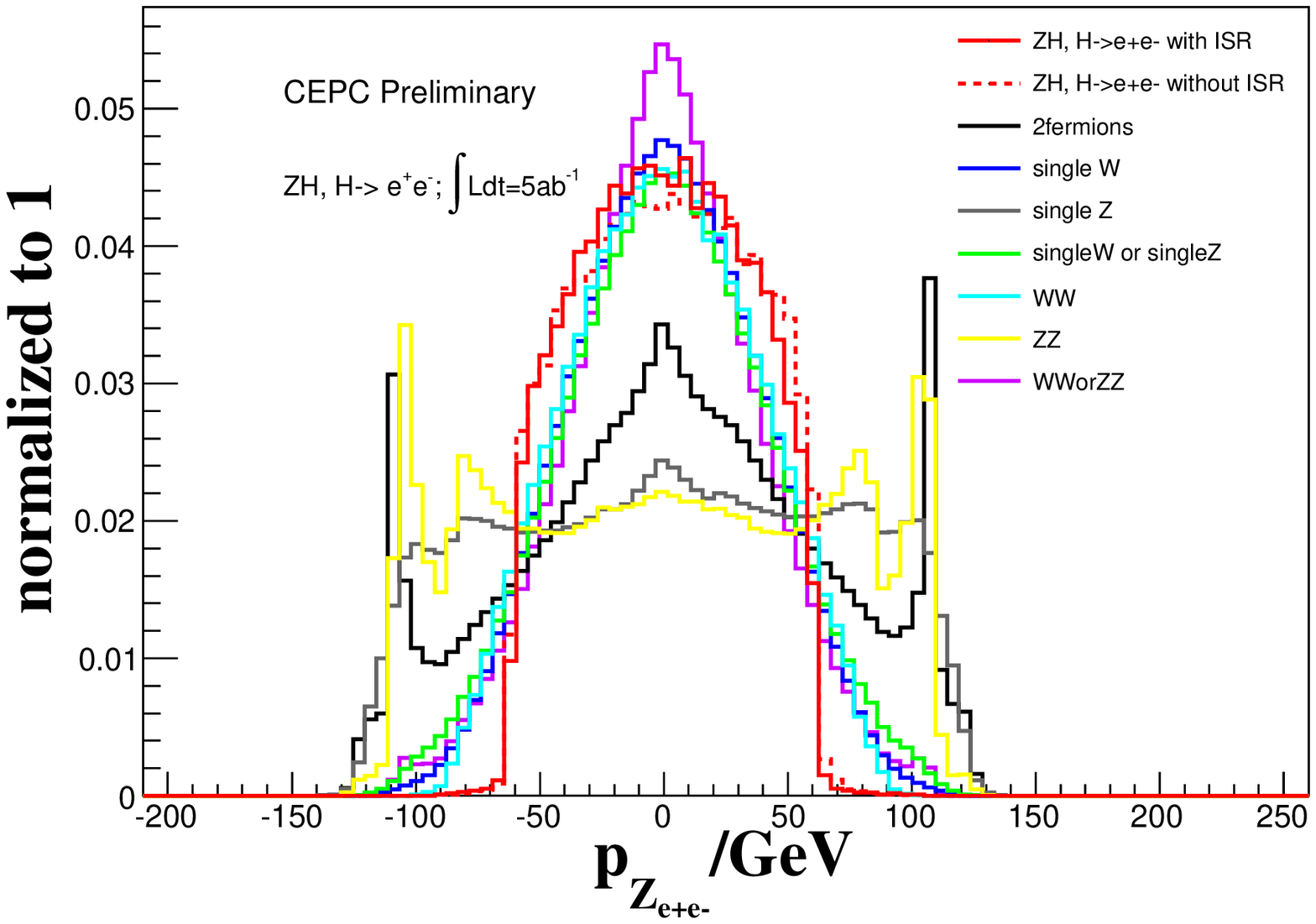}
  \includegraphics[width=0.4\textwidth]{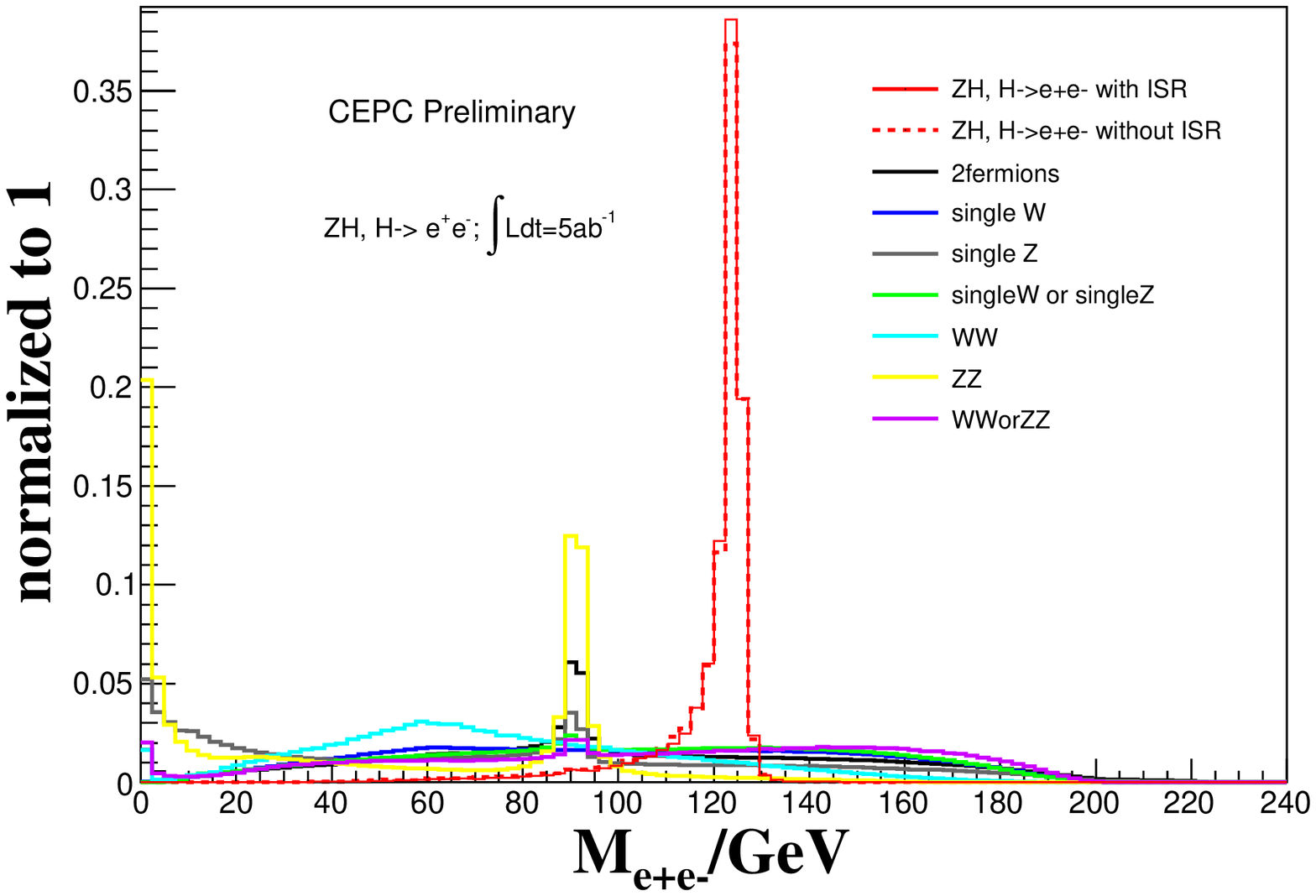}
  \includegraphics[width=0.4\textwidth]{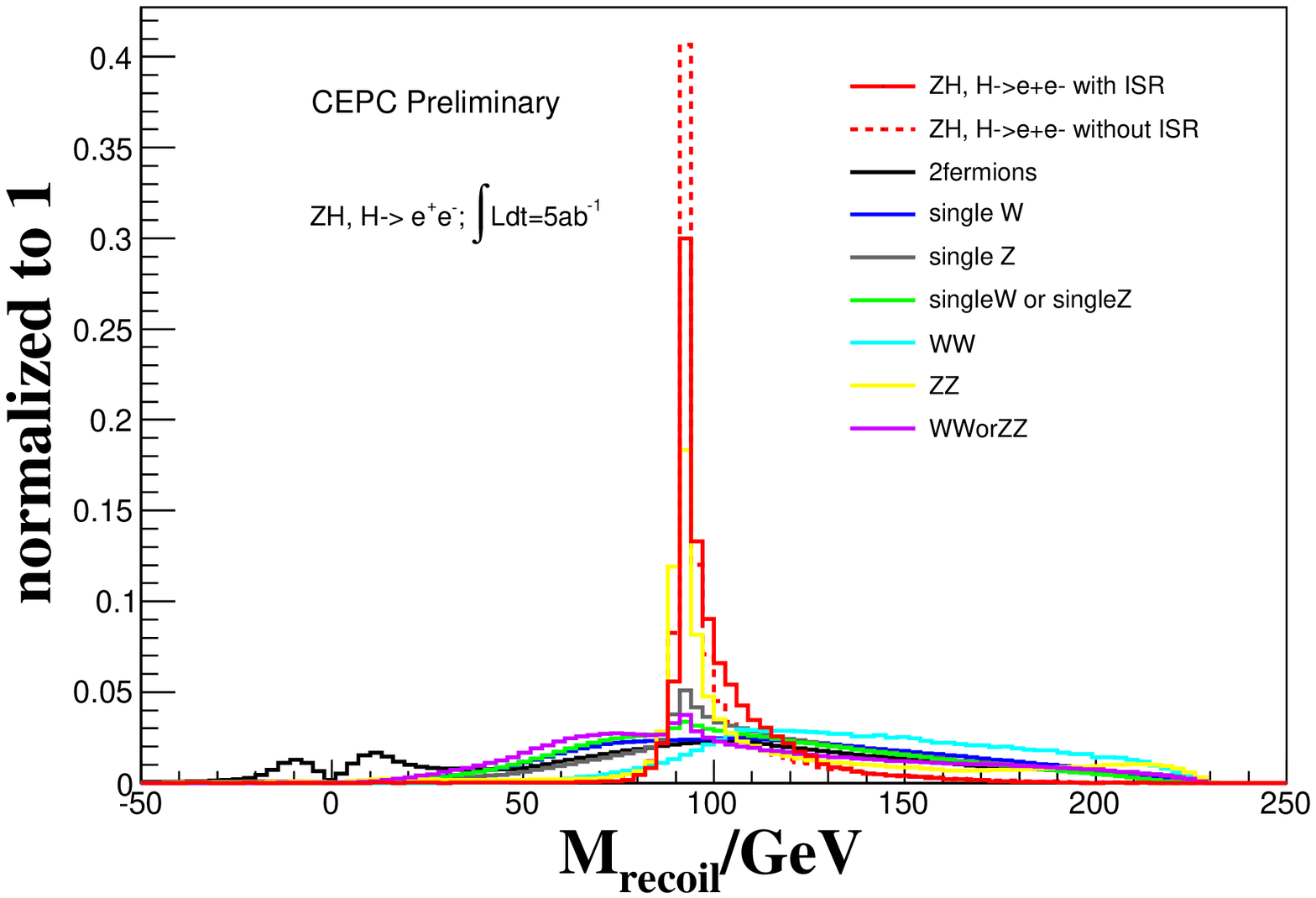}
  \caption{\label{fig:pt} Distributions of $p_{\rm{T}e^{+}e^{-}}$, $p_{\rm{Z}e^{+}e^{-}}$, $M_{e^{+}e^{-}}$ and $M_{recoil}$  for signals and backgrounds. Signals without ISR effect included are also superimposed for comparison.}
\end{figure}

\qquad As mentioned in Section~\ref{intr}, signal events can be extracted with recoil mass method without using the decay information of the Z boson decay. The detailed event selections are listed as following:  at least one pair of electrons with opposite charge is required, with final state radiation photon in included in the electron momenta.  The pair with invaraiant mass $M_{e^{+}e^{-}}$ closer to Higgs mass is selected in case of multi-combinations, and required then to satisfy $120<M_{e^{+}e^{-}}<130$\,GeV.  The recoil mass $M_{recoil}$ of $e^{+}e^{-}$ is required to be greater than 90 GeV and less than 93 GeV, to be consistent with the Z-boson hypothesis. Fig.~\ref{fig:pt} shows signal and backgrounds distributions on various kinematic variables, where signal without ISR effect included are also superimposed for comparison.

To suppress 2-fermions background, it is required that the difference between the two electrons' azimuth angles should satisfy $\Delta\phi\ <$ 166$^{\circ}$. In addition, to suppress background from 4-fermions background, the transverse momentum of electron pair and the scalar sum over Z-direction momentum, are required to $46<p_{\rm{T}e^{+}e^{-}}<93$\,GeV and $-42<p_{\rm{Z}e^{+}e^{-}}<41$\,GeV, which can efficiently cut away $ZZ$ and single $Z$ backgrounds, as shown in Fig.~\ref{fig:pt}. Finally, requirements are set on polar angle of each lepton particle,  $cos_{e^{+}}$ $\ge$ -0.07 and $cos_{e^{-}}$ $\le$ 0.14 , as the electrons from Higgs boson are more uniformly distributed as it is a scalar particle. The selections of each variable as mentioned above are determined by maximazing the significance $S/\sqrt{B}$, where S is the number of signal events passing all the selection criteria, and B is the number of the corresponding background events number.  

The cut chain table is shown in Table~\ref{tab:example1}. The background yields are scaled to 5000$fb^{-1}$~. The signal yields starts from 50K before any selection, and the final efficiency is about 7.1\%.

\begin{table}
  \begin{center}
    \begin{tabular}{lllllllllll}
    \hline \hline
      Category                                       &signal  &2fermions   &single ZorW   &single Z   &single W       \\
      \hline
      total                                          &50000   &418194802   &1259165       &7913405    &17190655                     \\
      $N_{e^{+}}$ $\ge$ 1, $N_{e^{-}}$ $\ge$ 1       &47418   &36822471    &978594        &3480494    &2260761                     \\
      120 GeV $<\ M_{e^{+}e^{-}}\ <$ 130 GeV             &34463   &1954192     &71193         &126094     &151950                   \\
      90 GeV $<\ M_{recoil}\ <$ 93 GeV           &12362   &61089       &3564          &6954       &7255                   \\
    46 GeV $<\ p_{\rm{T}e^{+}e^{-}}\ <$ 63 GeV       &8582   &6816        &1863          &1861       &3652             \\
    -42 GeV $<\ p_{\rm{Z}e^{+}e^{-}}\ <$ 41 GeV      &8511   &6372        &1783          &1750       &3468            \\
      $\Delta\phi\ <$ 166$^{\circ}$                  &7404   &5131        &1696          &1651       &3233                               \\
$cos_{e^{+}}$ $\ge$ -0.07, $cos_{e^{-}}$ $\le$ 0.14  &3564    &241         &86            &48         &161                \\
\hline
    \end{tabular}
\vspace{0.1cm} \\
    \begin{tabular}{lllllllllll}
    \hline  
      Category                                       &WW  &ZZ   &WWorZZ   &total background          \\
      \hline
      total                                          &49115769   &4967152   &21902983       &520543931                         \\
      $N_{e^{+}}$ $\ge$ 1, $N_{e^{-}}$ $\ge$ 1       &640839   &758732    &814608        &45756499                         \\
      120 GeV $<\ M_{e^{+}e^{-}}\ <$ 130 GeV             &26731   &7593     &55196         &2392949                        \\
      90 GeV $<\ M_{recoil}\ <$ 93 GeV           &1783   &1464       &2434          &84543                         \\
    46 GeV $<\ p_{\rm{T}e^{+}e^{-}}\ <$ 63 GeV       &868   &682        &1297          &17039                   \\
    -42 GeV $<\ p_{\rm{Z}e^{+}e^{-}}\ <$ 41 GeV      &837   &647        &1247          &16104                  \\
      $\Delta\phi\ >$ 166$^{\circ}$                  &702   &566        &1182          &14161                                     \\
$cos_{e^{+}}$ $\ge$ -0.07, $cos_{e^{-}}$ $\le$ 0.14  &20    &178         &70           &804                        \\
      \hline \hline
    \end{tabular}
  \caption[Cut-flow]{Yields for backgrounds and signals at the CEPC with $\sqrt{s}=250$ GeV and integrated luminosity of $5000\fbinv$.}
  \label{tab:example1}
  \end{center}
\end{table}

We have also exploited the Toolkit for Multivariate Analysis (TMVA)~\cite{TMVA} for further background rejection, where the  method of Boosted Decision Trees (BDT) is adopted and the
selected variables for TMVA input are those as mentioned above.  No significant improvement is found compared with the cut-based results, thus in this study, we provide only the latter.

\begin{figure}[!t]
  \centering
  \includegraphics[width=0.7\textwidth]{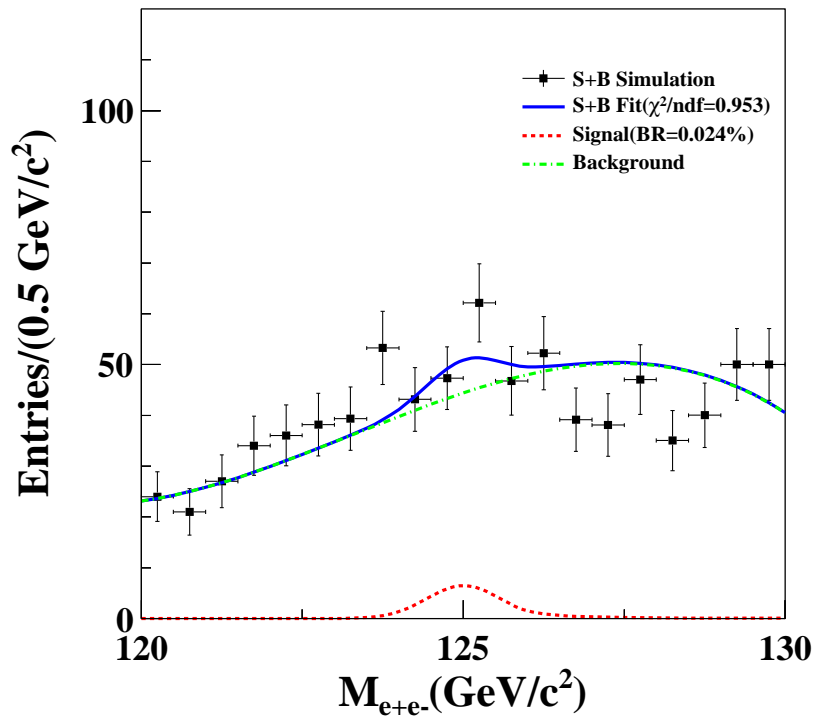}
  \caption{\label{fig:fit} The invariant mass spectrum of $e^{+}e^{-}$ in the inclusive analysis. The dots with error bars represent data from CEPC simulation. The solid (blue) line indicates the fit. The dashed (red)  shows the signal (assuming ${\cal B}(\Hboson \xrightarrow {}\ee)$=0.024\%) and the long-dashed (green) line is the  background.}
\end{figure}

After the event selections as mentioned above, we perform a $\mu S + B$ fit (with $\mu$ as the signal strength) on CEPC simulated data which is essential purely background as the SM predicted $\Hboson \xrightarrow {}\ee$~branch ratio is too low. As shown in Fig.~\ref{fig:fit}, an unbinned maximum likelihood fit is performed on $M_{e^{+}e^{-}}$ spectrum, in the region of 120 GeV to 130 GeV. The Higgs signal shape is described by a Crystal Ball function, while the background is represented by a second order Chebychev polynomial function, whose parameters are fixed to the values extracted from the background samples. By scanning over signal strength in the $\mu S + B$ fit, one can extract the dependence of negative log likelihood on it. The 95\% confidence level upper limit on $\Hboson \xrightarrow {}\ee$~branch ratio can then be decided to be  0.024\%.  This corresponds to a signal yield of around 20, while from Figure~\ref{fig:fit}, the background yield under the Higgs peak is near 200, and thus by naively couting, $S/\sqrt{B}\sim 1.4$ which supports the above result from the shape analysis.  Finally we mention that checks with different background modelling have also been done. With e.g. third order Chebychev polynomial function, the result improves a bit while the fit goodness gets worse.

\section{Summary and Conclusions}
\label{talk}
\qquad The CEPC is expected to play a crucial role in understanding Higgs boson properties. In this paper, a probe on $\Hboson \xrightarrow {}\ee$ at CEPC is investigated with full simulated Higgsstrahlung signal at 5 ab$^{\rm{-1}}$ integrated luminosity at 250 GeV center-of-mass energy. The upper limit at 95\% confidence level on the production cross section times branching fraction for $e^+e^- \rightarrow ZH$ with $\Hboson \xrightarrow {}\ee$ are found to be 0.051 fb. This corresponds to an upper limit on the branching fraction of 0.024\%. As a by-product, ISR effect has been implemented in \madgraph~ to generate the signal process.  Finally, we mention that with similar framework, measurements for $\Hboson \xrightarrow {}\mm$ together with $\tt$ at CEPC are being finalized~\cite{cepccdr}, which show similar or even improved accuracy compared with the results  for HL-LHC~\cite{ATLprojection,CMS:2013xfa}.

\acknowledgments
This work is supported in part by the National Natural Science Foundation of China, under Grants No. 11475190 and No. 11575005,  by the CAS Center for Excellence in Particle Physics (CCEPP), and by CAS Hundred Talent Program (Y3515540U1).

\appendix



\begin{thebibliography}{00}
\bibliographystyle{apsrev}
\bibliography{references}


\bibitem{ref:1}The ATLAS Collaboration, G. Aad et al., Phys. Lett. B, 2012, {\bf 716}: 1---29
\bibitem{ref:2}The CMS Collaboration, S. Chatrchyan et al., Phys. Lett. B, 2012, {\bf 716}: 30---61
\bibitem{cmshig} S. Chatrchyan et al (The CMS Collaboration), JHEP, {\bf 06}: 081 (2013)
\bibitem{lhcsub1} G. Aad et al (The ATLAS Collaboration), Phys. Lett. B, {\bf 726}: 88 (2013)
\bibitem{lhcsub2} G. Aad et al (The ATLAS Collaboration), Phys. Lett. B, {\bf 726}: 120 (2013)
\bibitem{lhcsub3} V. Khachatryan et al (The CMS Collaboration), Eur. Phys. J. C, {\bf 75}: 212 (2015)
\bibitem{lhcsub4} V. Khachatryan et al (The CMS Collaboration), Phys. Rev. D, {\bf 92}: 012004 (2015)
\bibitem{lhcsub5} G. Aad et al (The ATLAS Collaboration and CMS Collaboration), Phys. Rev. Lett., {\bf114}: 191803 (2015)
\bibitem{ref:3}CEPC-SppC Preliminary Conceptual Design Report: Physics and Detector, by the CEPC Study Group.
\bibitem{Chen:2016zpw} 
  Z.~Chen, Y.~Yang, M.~Ruan, D.~Wang, G.~Li, S.~Jin and Y.~Ban,
  Chin.\ Phys.\ C {\bf 41}, no. 2, 023003 (2017)
\bibitem{Altmannshofer:2015qra} 
  W.~Altmannshofer, J.~Brod and M.~Schmaltz,
  JHEP {\bf 1505}, 125 (2015)
  doi:10.1007/JHEP05(2015)125
  [arXiv:1503.04830 [hep-ph]].
\bibitem{Khachatryan:2014aep} 
  V.~Khachatryan {\it et al.} [CMS Collaboration],
  Phys.\ Lett.\ B {\bf 744}, 184 (2015)
\bibitem{fccee} Talk from David d'Enterria, https://indico.cern.ch/event/469576
\bibitem{ref:4}W. Kilian, T. Ohl, J. Reuter, Eur. Phys. J. C, 2011, {\bf 71}:1742.
\bibitem{ref:cepc_acc} CEPC Accelerator Preliminary Conceptual Design Report, by the CEPC Study Group.
\bibitem{ild1} The ILD concept group, arXiv:1006.3396
\bibitem{ild2} T. Behnke, J. Brau, P. Burrows et al, arXiv: 1306.6329
\bibitem{ILC} H. Baer, T. Barklow, K. Fujii et al, arXiv:1306.6352
\bibitem{ref:17}D. Bardina, M. Bilenkya, D. Lehnerc, A. Olchevskib and T. Riemann, Nucl. Phys. Proc. Suppl. B, 1994, {\bf 37}:148-157.
\bibitem{Mo:2015mza} 
  X.~Mo, G.~Li, M.~Q.~Ruan and X.~C.~Lou,
  Chin.\ Phys.\ C {\bf 40}, no. 3, 033001 (2016)
\bibitem{pythia}
  T.~Sjostrand, L.~Lonnblad, S.~Mrenna and P.~Z.~Skands,
  hep-ph/0308153
\bibitem{Mokka} Mora de Freitas, P. and Videau, H., LC-TOOL-2003-010
\bibitem{arbor} Manqi Ruan, arXiv: 1403.4784
\bibitem{TMVA} P. Speckmayer, A. Hocker, J. Stelzer, and H. Voss, J.Phys.Conf.Ser. 219 (2010) 032057
\bibitem{cepccdr} CEPC-SppC Conceptual Design Report: Physics and Detector, by the CEPC
Study Group, to appear soon
\bibitem{ATLprojection}
[ATLAS Collaboration], ATL-PHYS-PUB-2013-014
\bibitem{CMS:2013xfa}
  [CMS Collaboration],
  arXiv:1307.7135 [hep-ex].

\end{thebibliography}
\end{document}